\documentclass[prl,twocolumn,superscriptaddress,showpacs]{revtex4}
\usepackage{amsmath,amssymb}
\usepackage[utf8x]{inputenc}
\usepackage[english]{babel}
\usepackage{graphicx,epsfig}
\usepackage{color}
\usepackage{hyperref}
\usepackage{txfonts}

\global\long\def\v#1{\vec{#1}}
\global\long\def\a{\hat b}
\global\long\def\ad{\hat b^\dagger}
 \global\long\def\abs#1{\left|#1\right|}
\global\long\def\ket#1{\left| #1\right\rangle }
 \global\long\def\bra#1{\left\langle #1 \right|}
  \global\long\def\av#1{\left\langle #1 \right\rangle }
  
\newcommand{\ph}{\text{ph}}
\newcommand{\IB}{\text{IB}}
\newcommand{\p}{\text{p}}
\global\long\def\tens#1{\boldsymbol{#1}}
\global\long\def\mI{\mathcal{M}^{-1}}

\begin{document}

\title{Polaronic mass renormalization of impurities in BEC: correlated Gaussian wavefunction approach}

\author{Yulia E. Shchadilova}
\affiliation{Russian Quantum Center, Skolkovo 143025, Russia}
\email{yes@rqc.ru}
\affiliation{Department of Physics, Harvard University, Cambridge, Massachusetts 02138, USA}

\author{Fabian Grusdt}
\affiliation{Department of Physics and Research Center OPTIMAS, University of Kaiserslautern, Germany}
\affiliation{Graduate School Materials Science in Mainz, Gottlieb-Daimler-Strasse 47, 67663 Kaiserslautern, Germany}
\affiliation{Department of Physics, Harvard University, Cambridge, Massachusetts 02138, USA}

\author{Alexey N. Rubtsov}
\affiliation{Russian Quantum Center, Skolkovo 143025, Russia}
\affiliation{Department of Physics, Moscow State University, 119991 Moscow, Russia}

\author{Eugene Demler}
\affiliation{Department of Physics, Harvard University, Cambridge, Massachusetts 02138, USA}
\date{\today}

\begin{abstract}
We propose a class of variational Gaussian wavefunctions to describe Fr\"ohlich polarons at finite momenta.
Our wavefunctions give polaron energies that are in excellent agreement with the existing Monte Carlo results for a broad range of interactions.
We calculate the effective mass of polarons and find smooth crossover between weak and intermediate impurity-bosons coupling.
Effective masses that we obtain are considerably larger than those predicted by the mean-field method.
A novel prediction based on our variational wavefunctions is a special pattern of correlations between host atoms that can be measured in time-of-flight experiments.
We discuss atomic mixtures in systems of ultracold atoms in which our results can be tested with current experimental technology. 
\end{abstract}

\maketitle

\begin{figure}[t]
\centering
\includegraphics[width=1\columnwidth]{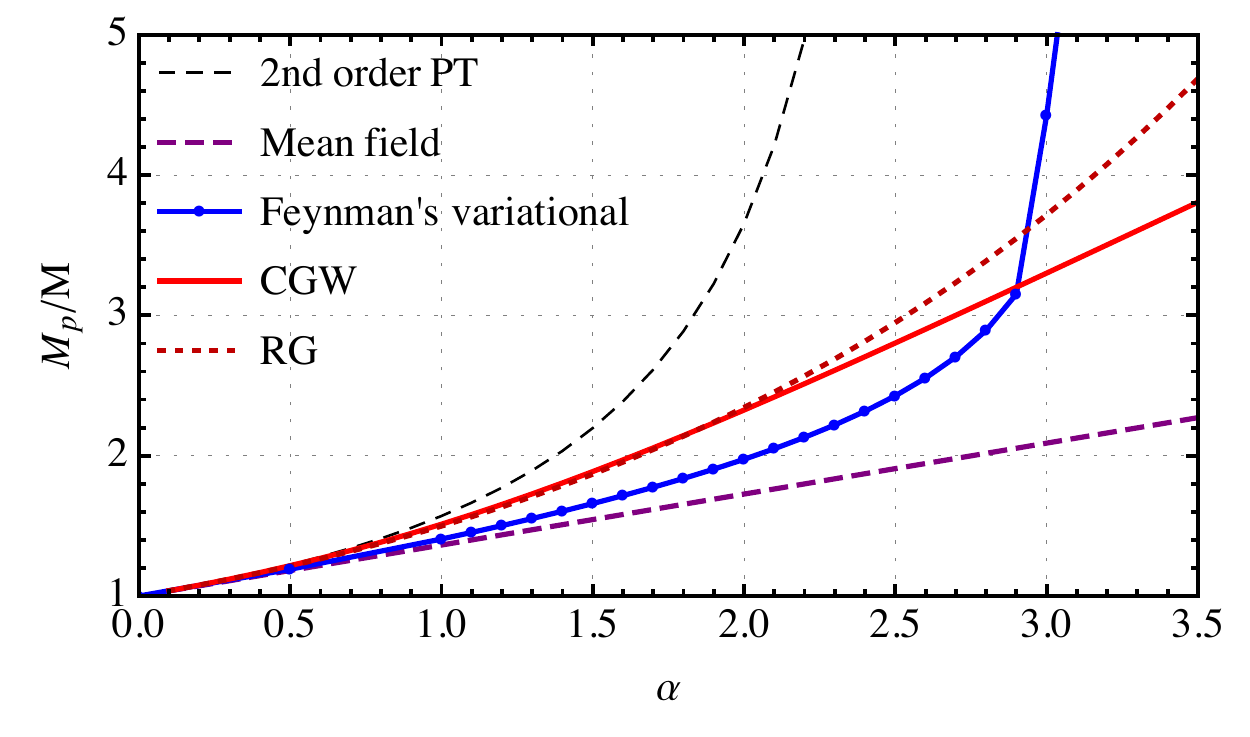}
\caption{
The polaron effective mass computed using 
different approaches as a function of the dimensionless interaction parameter $\alpha$ defined by Eq.~(\ref{eq:alpha}). 
Our result (CGWs) is compared with 
Feynman's variational method
\cite{Casteels2012pra}, mean-field ~\cite{aditya}, and Renormalization Group approach (RG)~\cite{RG}.
We used parameters corresponding to $^6$Li impurities in $^{23}$Na BEC for comparison with Refs.
~\cite{Tempere2009,Vlietinck2014}.}
\label{fig:Mass_theory_comparison}
\end{figure}

Renormalization of particle masses due to their interaction with the environment is a ubiquitous phenomenon in 
physics. In the standard model of high energy physics elementary particles acquire a mass
through interaction with the Higgs field~\cite{Weinberg_book}. 
In solid state systems heavy fermion materials exhibit renormalization of electron masses of up to two orders 
of magnitude  due to interaction of electrons with localized spins~\cite{review_Qi.S}. 
Complete localization of quantum degrees of freedom caused by interaction with the environment has been discussed in
spin-bath models~\cite{Leggett_RMP,Silbey1984} and quantum Josephson junctions~\cite{Schon1990,chakravarty,Penttila1999}.
Surprisingly one of the first systems in which strong mass renormalization due to particle-bath
interaction has been predicted, the so-called polaron model introduced by Landau in 1933~\cite{Landau,Pekar},
remains a subject of considerable debate. This model describes interaction of a quantum particle with a bosonic bath, 
such as an electron interacting with phonons in a crystal (see Refs.~\cite{Mitra1987,Devreese2009,Emin2013} for reviews).  
While the limiting cases
of weak and strong coupling can be analyzed using controlled perturbtative expansions 
(see Refs.~\cite{Frohlich1954,Lee1953} for weak coupling analysis and Refs.~\cite{Pekar,Allcock1963} for strong coupling expansion), the intermediate coupling regime remains poorly understood
with the effective mass of the polaron being the most contentious issue~\cite{Fisher1986}.
For example, considerable disagreement between different approximations for the effective mass of polarons in BEC has been reported in the literature  (see Fig.\ref{fig:Mass_theory_comparison}).
Perturbative expansion for small interaction strength suggest a divergence of the effective mass beyond a certain interaction strength, indicating  localization of the impurity particle~\cite{Mahan}. Variational method based on the Feynman path integral  approach does not have 
a phase transition but  exhibits a sharp crossover in the effective mass~\cite{Tempere2009,Casteels2012pra}. By contrast mean-field approach to the problem shows only a gradual evolution of the effective mass~\cite{aditya}. 
Recent experimental progress
in the field of ultracold atoms brought new interest in the study of impurity problems. Feshbach resonances made it
possible to realize both Fermi~\cite{zwierlein,grimm,kohl,Nascimbene2009} and Bose polarons
~\cite{inguscio,Chikkatur2000,Palzer2009,Zipkes2010,Heinze2011,Spethmann2012,Chen2014} with tunable interactions between impurity and host atoms
and the rich toolbox of atomic physics has been used to study their properties including the effective mass
~\cite{Nascimbene2009,zwierlein}. 

In this paper we show that an analytical class of wavefunctions based on the  correlated Gaussian ansatz can describe Fr\"ohlich polarons at finite momentum for a wide range of parameters 
(see Refs.~\cite{Tulub1961,Altanhan1993,Kandemir1994} for earlier works). 
%We use our model to calculate the effective mass of polarons in the specific case of polarons realized with impurities in BEC of ultracold atoms. 
Fr\"ohlich type Hamiltonians can be used to describe several different families of physical systems including 
electrons interacting with lattice phonons in polar~\cite{Landau,Pekar,Feynman1955,Tulub1961,Altanhan1993,Kandemir1994,Mishchenko2000,Prokof'ev2008}, organic
~\cite{Vukmirovic2012}, and piezoelectic
~\cite{Mahan1964,Volovik1973,Volovik1975} semiconductors,
magnetic polarons in strongly correlated electron systems
~\cite{Nagaev1975,Trugman1988}, 
$^3$He atoms in superfluid $^4$He
~\cite{Bardeen1967},
and impurity atoms in Bose-Einstein condensates
~\cite{Cucchietti2006,Klein2007,Tempere2009,Privitera2010,Novikov2010,
Casteels2011pra,Casteels2011lp,Casteels2012pra,Rath2013,Cucchietti2013,Casteels2013,aditya,Li2014,Vlietinck2014,Kain2014,RG,Christensen2015,Levinsen2015}.
In this paper we focus on the BEC polarons realized with ultracold atoms since these systems have highly tunable paramaters and should allow most
detailed comparison with our theoretical analysis. However our method can be easily adapted to other systems and generalized to dynamics.

\begin{figure}[t]
\includegraphics[width=1\columnwidth]{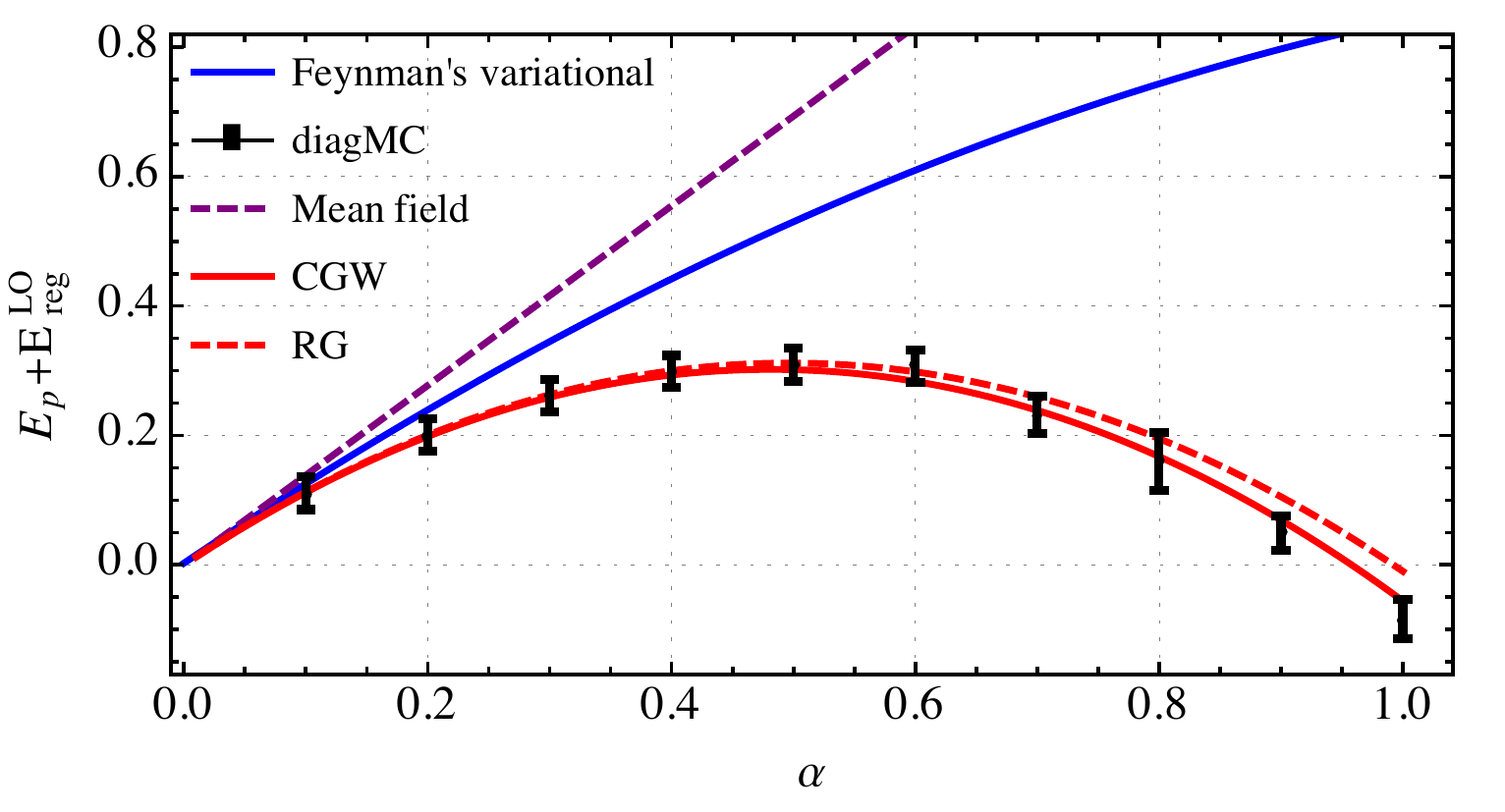}
\caption{
Polaron energy {$E_p + E_\text{reg}^\text{LO}$} for static $^6$Li impurity ($P=0$) in $^{23}$Na BEC predicted by different theoretical approaches
as a function of the dimensionless coupling constant $\alpha $.
Our result (CGWs) is compared with MC calculations~\cite{Vlietinck2014}, Feynman's variational method~\cite{Tempere2009}, mean-field
~\cite{aditya}, and RG~\cite{RG}.
Note that polaron energy  {has a sub-leading UV divergence}. We choose the cut-off $2\cdot10^3 \xi^{-1}$ as in Refs.~\cite{Tempere2009,Vlietinck2014}.
 }
\label{fig:Energy_theory_comparison} 
\end{figure}

\begin{figure}[t]
\centering
\includegraphics[width=1\columnwidth]{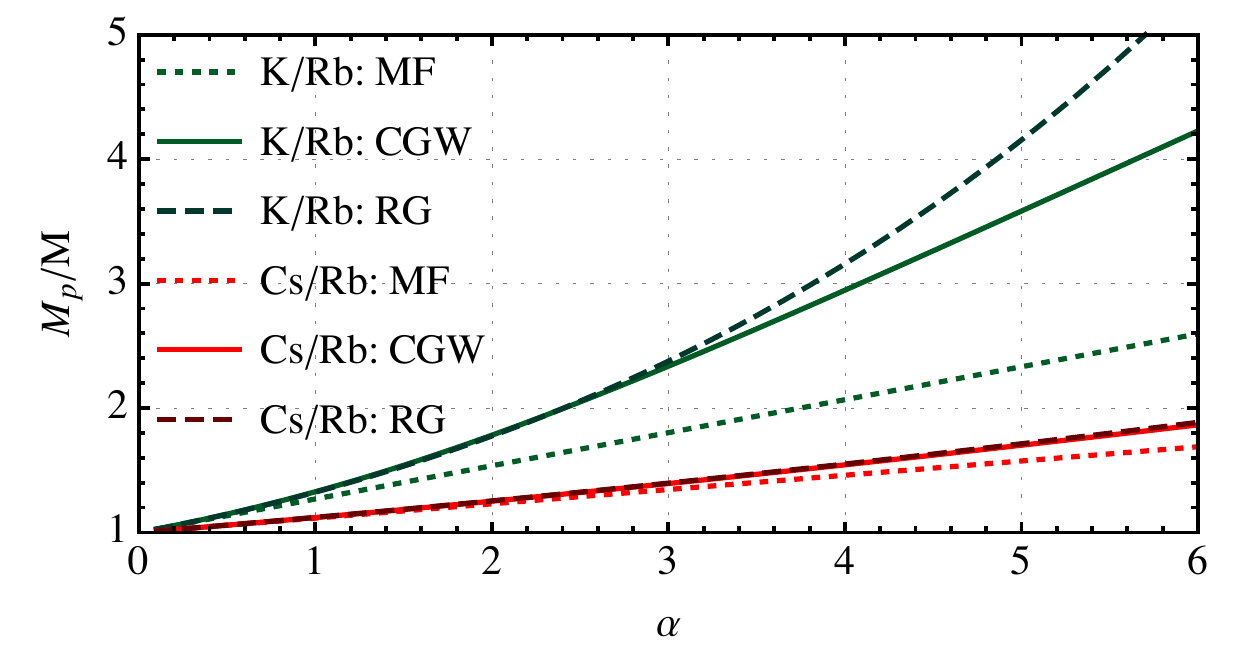}
\caption{
Polaron mass for $^{41}$K and $^{133}$Cs impurities in BEC of $^{87}$Rb atoms 
(in units of bare impurity mass $M$).
Increase of the interaction strength $\alpha$ between impurity atom and BEC
enhances quantum fluctuations and  results in stronger renormalization of $M_\p$.
}
\label{fig:1}
\end{figure}

\begin{figure}[t]
\includegraphics[width=0.83\columnwidth]{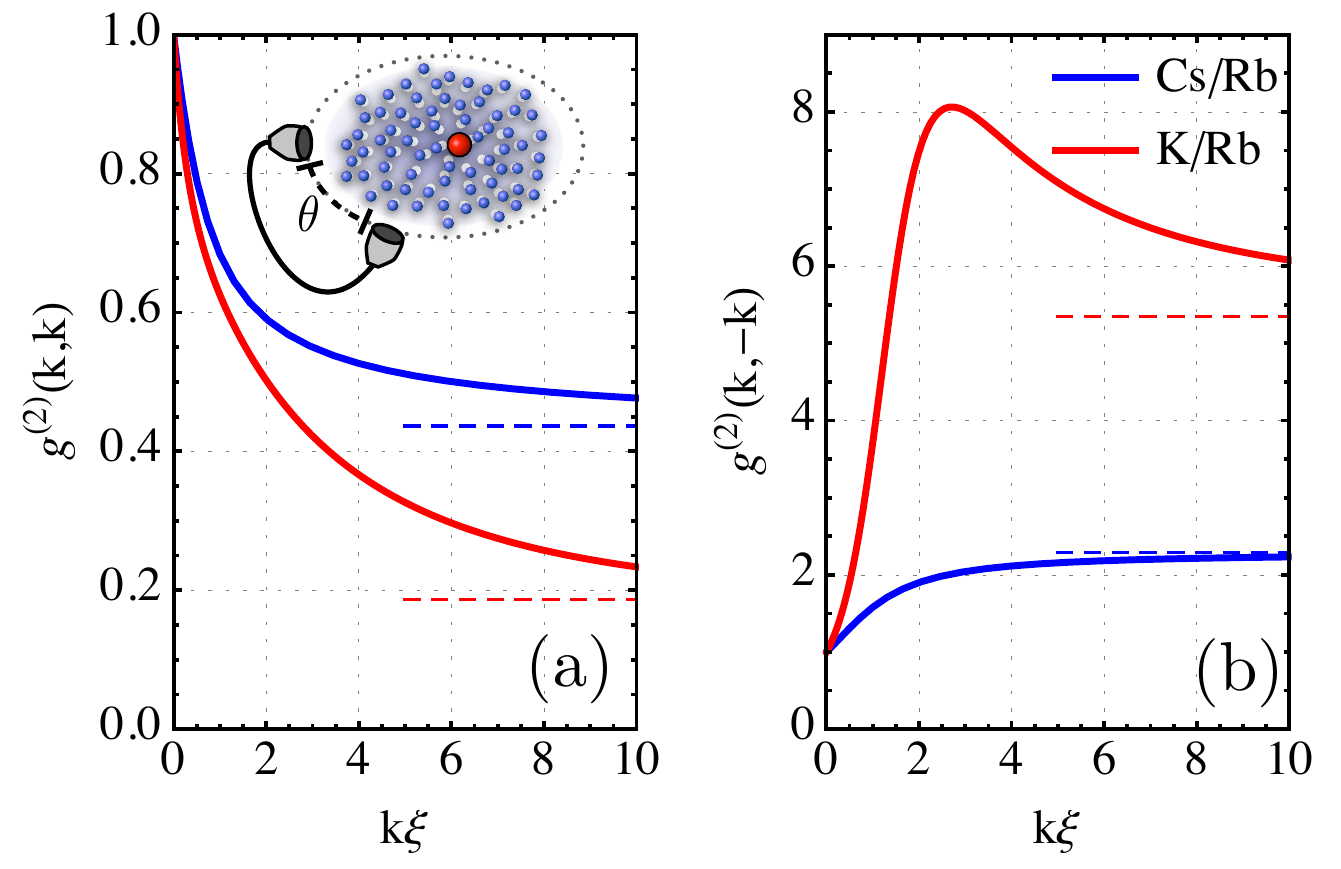} 
\caption{Inset in the panel (a) shows the typical experimental setup for measuring noise correlation functions
$g^{(2)}(k,k')$ (see Eq.(\ref{Eq:g2})). TOF measurement should be performed with two detectors placed at relative angle $\theta$ 
between two directions of measurement.
Panels (a) and (b) show noise correlations for (a) $\theta= 0$ and (b)  for $\theta=\pi$ for
$^{41}$K/$^{87}$Rb and $^{133}$Cs/$^{87}$Rb systems at $\alpha=4$. In the case $\theta= 0$
BEC atoms show antibunching. In the case of $\theta=\pi$ we find atom bunching. 
For the $^{41}$K/$^{87}$Rb mixture this bunching has a peak at $k/\xi=3$. Large momentum asymptotics can be computed analytacally and are shown
with dashed lines.
 }
\label{fig:2}
\end{figure}

The essence of our approach is an extension of the earlier mean-field variational wavefunction 
(after performing Lee-Low-Pines transformation on the Hamiltonian, 
see~\cite{Aleksandrov_polaron_book} and discussion below) to Gaussian wavefunctions that include
entanglement between different phonon modes. 
The explicit form of these wavefunctions is given in equation (\ref{CGW}) and from now on we will refer to them as correlated Gaussian wavefunctions (CGW)
~\cite{Mitroy2013}. We demonstrate that CGWs show excellent agreement with the
Monte Carlo (MC) results for the polaron binding energy at zero momentum  for the range of interactions relevant to ultracold atoms: from weak to intermediate coupling (see Fig.\ref{fig:Energy_theory_comparison}). Compared to the mean-field solution the number of variational parameters in our approach increases only by a factor of three, which keeps the number of self-consistent equations reasonably small. 
In the context of BEC polarons MC method has only been used to calculate polaron binding energies at zero momentum, which makes it impossible to obtain the effective mass (for other types of polarons however effective mass analysis based on MC calculations has been done, see e.g.~\cite{Mishchenko2000, Prokof'ev2008}.)
CGW analysis at finite polaron momentum does not introduce additional complications which allows us to calculate the effective mass of polarons. The key ingredient of our new class of wavefunctions is the appearance of additional correlations  for the host atoms introduced through their interaction with mobile impurities. These correlations can be observed in the time of flight experiments and should provide a quantitative test of our variational wavefunction. Intermediate coupling regime of polarons should be accessible, e.g., with  $^{41}$K  or $^{133}$Cs atoms  in $^{87}$Rb BEC, both of which have interspecies Feshbach resonances that can  be used to tune the impurity-boson interaction. In Fig.\ref{fig:1} we present our predictions for polaron masses for the two cases and in Fig.\ref{fig:2} we present predictions for the BEC atoms correlations.

%============================================================

{\it Fr\"ohlich model for the impurity atom interacting with BEC}. 
We use Bogoliubov model to describe BEC of the host atoms
and limit ourselves to small deviations of the BEC density from the homogeneous case. 
In this case interaction of the impurity with phonons of the BEC can be described using the Fr\"ohlich Hamiltonian
~\cite{Cucchietti2006,Sacha,Tempere2009}
\begin{equation}
\label{H_Frohlich}
\hat{\cal H} = \frac{\hat{\v p}^2}{2 M} +
\sum_{k}  V_{k}  \left( \a_{k} + \ad_{-k}  \right)  e^{i \v k \hat{\v r}} 
 + \sum_{k}\omega_{k} \ad_{k} \a_{k}
\end{equation}
Here $\hat{\v p}$ and $\hat{\v r}$ are momentum and position operators of the impurity atom with mass $M$, 
$\ad_{k}$ is the annihilation operator of the Bogoliubov phonon excitation with momentum $\v{k}$, 
 $\omega_k= ck \sqrt{1+(\xi k)^2/2}$ is the Bogoliubov mode dispersion, with $c$ being the sound velocity and
$\xi$ is the coherence length of the condensate. The impurity-phonon interaction 
strength  is given by
$V_k =  g_{\rm IB} \sqrt{n_0 \mathcal{V}^{-1}  \xi k} \left(2+\left(\xi k\right )^2\right)^{-1/4}  $,
where $n_0$ is the BEC density,
$g_{\rm IB}$ denotes the interaction strength between the impurity atom and host 
atoms with mass $m$,  and $\mathcal{V}$ is the volume of the system. From now on we will set 
$\mathcal{V}=1$ in the rest of the paper.
In the first-order Born approximation this interaction strength can be related to the impurity/BEC atom scattering length
via $g_\IB = 2 \pi a_\IB \left( m^{-1}+M^{-1}\right)$. 
Conditions on the applicability of the Fr\"ohlich model for describing impurity-boson systems are discussed below
when we consider experimental systems.

To utilize translational symmetry of the Fr\"ohlich Hamiltonian (\ref{H_Frohlich}) we apply the 
Lee-Low-Pines (LLP) transformation~\cite{Lee1953} $\hat{U}_{\text{LLP}}= e^{i\hat S}$, $\hat S= \hat{\vec{r}} \sum_k \vec{k} \ad_k \a_k$.
The transformed Hamiltonian $\hat{\cal H}_{\rm LLP} = \hat{U}_{\text{LLP}} \hat{\mathcal{H}} \hat{U}_{\text{LLP}}^\dagger$ is
\begin{eqnarray}
\label{H_LLP}
&&\hat{\cal H}_{\rm LLP}= \frac{1}{2M}\left( \v P-\sum_k \v k \ad_{k} \a_{k}\right)^2+\nonumber \\
&& \hspace{2cm} \sum_k  \omega_{ k}  \ad_{ k}  \a_{ k}  
+\sum_k V_{ k} \left(  \a_{ k} +  \ad_{- k}\right)
\end{eqnarray}
Here $\v P$ is a conserved total momentum of the system which can be treated as a c-number. Equation (\ref{H_LLP}) 
no longer has degrees of freedom corresponding to the impurity: they were integrated out using
conservation of the total momentum. This generated an interaction term between phonon modes which is proportional to $1/M$. 
The appearance of the phonon-phonon interaction can 
be understood as exchange of momentum between phonons via the impurity.

%============================================================

{\it Mean-field solution}. To motivate the mean-field solution we first discuss the limit of infinitely heavy impurity, $M \rightarrow \infty$. In this case
interactions between phonon modes in eq. (\ref{H_LLP})
vanish and the Hamiltonian can be transformed to the canonical form using the displacement transformation 
$\hat D(\{\beta^0_k\}) = \exp \left (\sum_k\beta^0_k\ad_{k} -{\rm H.c.}\right)$ with $\beta^0_k =- V_k/\omega_k$.
Then the ground state is given by a coherent
state $\hat D(\{\beta^0_k\}) | 0 \rangle $, where $|0\rangle$ is the phonon vacuum. Note the key feature of this solution: it factorizes into a product of wavefunctions for different $k$-modes. Now we can generalize this result to the 
interacting case at finite $M$. The mean-field approach to polarons
assumes a similar structure of the polaron wavefunction even in the interacting case
of finite impurity mass~\cite{Alexandrov1995}. In this method  a product of coherent states for different phonon modes
is taken as a variational ansatz
\begin{eqnarray}
|MF \rangle = \hat D(\{\beta_k\}) | 0 \rangle
\label{mean_field}
\end{eqnarray}
and coefficients $\beta_k$ are determined from minimizing the energy
 $\langle  \hat{\cal H}_{\rm LLP} \rangle_\text{MF}$. 
 Straightforward calculation~\cite{aditya} gives
$\beta_{ k}= -V_{ k}/\Omega_{ k}$, where the renormalized dispersion $\Omega_k$ is given by
\begin{eqnarray}
\label{Omega_k}
\Omega_{ k}&=&\omega_{ k}+\frac{k^2}{2M}-\frac{\v k}{M}\left( \v P -\v P_\mathrm{ph}\right)
\end{eqnarray}
The parameter $\vec{P}_\mathrm{ph}$ describes the part of the total polaron momentum which is carried by the phonon cloud,
and in the mean-field approximation reads $\v P_\mathrm{ph}=\sum_q \v q\abs{\beta_{q}}^2$.
A major limitation of the mean-field state is that it does not include correlations between different phonon modes since the wavefunction factorizes into a product of wavefunctions for individual $k$s. Different modes affect each other only through the self-consistency condition 
on $\beta_k$. 

%============================================================

{\it Correlated Gaussian wavefuction}.
To account for correlations between different phonon modes in the polaron problem we introduce a Gaussian wavefunction
\begin{eqnarray}
\ket{\rm CGW}=\hat D(\lbrace\beta\rbrace) \hat S(\lbrace Q\rbrace) \ket{0}
\label{CGW}
\end{eqnarray}
where $\hat S(\lbrace Q \rbrace)=\exp( \frac{1}{2} \sum_{k,k'} Q_{k k'} \ad_{k} \ad_{k'} -H.c. )$. 
Variational wavefunction of this type have been suggested before
~\cite{Tulub1961,Altanhan1993,Kandemir1994}
but full optimization of the wavefunctions with respect to both $\beta$ and $Q$ was considered 
computationally impossible. For example, in Ref.~\cite{Tulub1961} energy was minimized with respect to
the boson displacement part $\lbrace\beta_k\rbrace$ and the Gaussian part was used to diagonalize 
Hamiltonian, where terms of the order higher than two were truncated. One of the key results of this paper is development of a
new approach finding the optimal values of $\beta$ and $Q$, which makes variational functions~\eqref{CGW} a powerful new tool 
for studying many-body systems of interacting bosons.

A convenient way of understanding
this ansatz is to interpret it as a generalized Bogoliubov transformation with 
\begin{multline}
\hat S^\dag(\lbrace Q \rbrace)  \hat D^\dag(\lbrace\beta\rbrace) 
\a_k \hat  D(\lbrace\beta\rbrace) \hat S(\lbrace Q\rbrace) = \\
\beta_k + \sum_{k'} \left[ \cosh Q\right]_{kk'} \a_{k'} +
\sum_{k'} \left[ \sinh Q\right]_{kk'} \ad_{k'} 
\label{Bogoliubov_transform}
\end{multline}
A new feature of wavefunction (\ref{CGW}) is that expectation
values of boson creation and annihilation operators no longer factorize. We have $\av{\a_k}=\beta_k$, 
$\av{\a \cdot \a} =\beta_k\cdot \beta_{k'} + (\cosh Q\sinh Q)_{kk'}$, and 
$\av{\ad \cdot  \a}= \beta_k \cdot   \beta_{k'} + (\sinh^2 Q)_{kk'}$. 
All higher order expectation values of $b_k$ and $b_k^\dagger$ operators can be computed using Wick's theorem. 
Variational parameters $Q_{kk'}$ and $\beta_k$ should be determined
by minimizing the energy. 

Explicit expression for the expectation value of $\hat {\cal H}_{\rm LLP}$ in state (\ref{CGW})
is given in Supplementary Materials (SM). In the regime of interest for cold atoms systems it is sufficient to expand the hyperbolic functions in 
(\ref{Bogoliubov_transform}) up to second order in matrices $Q_{kk'}$.
We find
\begin{multline}\label{eq:AvEnApprox}
\langle   \hat {\cal H}_{LLP}  \rangle_\text{CGW} =\frac{\vec{P}^2-\vec{P}^2_\text{ph}}{2M}+
\sum_k~\left( 2V_{k}\beta_k+\Omega_k \left( \beta_k^2+\sum_{k'}Q_{kk'}^2\right)\right) \\
\sum_{kk'}\frac{\vec{k} \vec{k'}}{M}Q_{kk'}^2+
\sum_{kk'}\frac{\vec{k}\vec{k'}}{M}\beta_k \beta_{k'}\left( Q_{kk'}+\sum_q Q_{kq}Q_{qk'}\right) 
\end{multline}
In this approximation the momentum of phonon cloud is defined as $\vec{P}_\mathrm{ph}=\sum_{k}\v k \beta_k^2+\sum_{kk'}\v k (Q_{kk'})^2$,
where $Q_{kk'}^2$ is the square of the matrix element. Minimizing expression with respect to $Q_{kk'}$ we obtain equations
\begin{multline} \label{eq:Q}
\left(\Omega_k +\frac{\vec{k}\vec{k'}}{M}+\Omega_{k'}\right)Q_{kk'}+
\frac{\vec{k}\vec{k'}}{M}\beta_k \beta_{k'}+\\
\sum_q \frac{\vec{q}}{M}\beta_q \left(\vec{k'} \beta_{k'}Q_{kq}+
\vec{k}\beta_{k}Q_{qk'}\right)=0
\end{multline}
where $\Omega_k$ is still given by equation (\ref{Omega_k}). At first sight this integral equation on the matrix $Q_{kk'}$ appears quite challenging. 
Fortunately, it can be reduced to a much simpler vector equation
by introducing $\vec{F}_k = - \beta_k^{-1} \sum_q \beta_q Q_{kq} \vec{q}$.
Then equation (\ref{eq:Q}) is equivalent to 
\begin{eqnarray}\label{F:eq}
\v F_k=\frac{1}{M}\sum_{k'} \frac{\beta_{k'}^2\v k'  }{\Omega_k+\frac{\v k \v k'}{2M}+\Omega_{k'}} \left(  \v k \v k' - \v F_k \v k' - \v k\v {F}_{k'}\right).
\end{eqnarray}
Minimization of  (\ref{eq:AvEnApprox})
with respect to $\beta_k$ gives
\begin{equation}\label{eq:a1}
V_{k}+\beta_k\Omega_k-\frac{ \v k}{M}\sum_q~\frac{\beta_q}{M }Q_{qk} \left( \v F_q -\v q\right)=0
\end{equation}
Eqs. (\ref{F:eq}) and (\ref{eq:a1}) can now be solved numerically. Details of the derivation of these equations can be found in SM.

To benchmark the approach we compare the ground state energy of the Fr\"ohlich Hamiltonian 
$E_p=\av{\hat{\mathcal{H}}_\mathrm{LLP}}$ for parameters corresponding to the system of Li impurity in Na BEC 
with other known theoretical results in Fig.\ref{fig:Energy_theory_comparison} ( {the comparison in strong coupling limit is provided in SM}).  
To make such comparison quantitative we regularize 
the leading order UV divergence of the polaron energy adding $E_\text{reg}^\text{LO} = 4 a_\text{IB}^2 n_0 (1+m/M) \Lambda $~\cite{Tempere2009,aditya}, where $\Lambda$ is the UV cut-off.
Our approach shows excellent agreement with the MC approach and drastically improves the mean-field solution.

%============================================================

{\it Effective mass of the polaron} can be obtained by
taking the second derivative of the polaron energy with respect to the total momentum. In practice this is not the most convenient way of computing it because the polaron energy for the Fr\"ohlich Hamiltonian (\ref{H_Frohlich})
has UV divergencies. To regularize these divergencies we should not only use a full cut-off dependent relation between the microscopic interaction $g_{IB}$ and the scattering length $a_{IB}$~\cite{Tempere2009,Rath2013,aditya}, but also consider renormalization
of the impurity mass that enters the mean-field part of the impurity interaction with the condensate. 
Detailed discussion of these divergences is presented in 
Ref.~\cite{RG}. One can however circumvent dealing with UV divergences if we use the 
following argument: When analyzing the variational wavefunction (\ref{CGW}) 
we can calculate momentum of the polaron carried by the impurity $\vec{P}_{\rm imp}=\vec{P}-\vec{P}_{\rm ph}$. The velocity 
of the impurity, ${P}_{\rm imp}/M$, should coincide with the velocity of the polaron, $ P/M_p$.
Thus we find
\begin{equation}
\frac{M}{M_p} = 1-\frac{P_{\ph}}{P}.
\end{equation}
For the comparison with other theoretical results we show the polaron mass for the $^6$Li impurities in $^{23}$Na BEC in Fig.\ref{fig:Mass_theory_comparison}.
In contrast with Feynman's variational approach the polaron mass calculated with CGWs shows smooth crossover from the regime of 
weak to intermediate coupling.

%============================================================

{\it Relevant experimental systems and results}. When selecting atomic mixtures for testing our theoretical analysis one needs to consider two factors. 
Firstly, a Feshbach resonance between the impurity and BEC atoms is required, that can be used to tune the interaction strength. 
It is common to describe the strength of this interaction using a dimensionless parameter 
\begin{equation}\label{eq:alpha}
\alpha = 8 \pi n_0 a_{IB}^2 \xi.
\end{equation}
Intermediate coupling regime requires $\alpha$ to be of the order of one.
Secondly, applicability of the Fr\"ohlich Hamiltonian relies on the condition that the condensate density depletion caused by the
impurity is smaller than the density of the condensate itself. 
This allows us to restrict ourselves to linear terms in Bogoliubov operators in the Fr\"ohlich Hamiltonian (\ref{H_Frohlich}) and gives rise to the condition
%This condition underlies using only linear terms in Bogoliubov operators in the BEC density operator in the Hamiltonian (\ref{H_Frohlich}). This gives a condition 
$ | g_{IB}|\ll 4c \xi^2$~\cite{Bruderer2007}. We find that both conditions can be satisfied for $~^{41}\text{K}$ impurities in  $~^{87}\text{Rb}$ BEC~\cite{Catani2008,inguscio} and $~^{133}\text{Cs}$ impurities in $~^{87}\text{Rb}$ BEC~\cite{mccarron2011dual,Spethmann2012}. They correspond to the cases of  moderately light impurities with  $M/m_B=0.46$ for
$~^{41}\text{K}$/${}^{87}\text{Rb}$ and  $M/m_B=1.53$ for $~^{133}\text{Cs}$/${}^{87}\text{Rb}$. Fig.\ref{fig:1}
shows our predictions for the effective mass of polarons in these two systems as a function of the impurity-boson interaction strength. 
We expect that mass enhancements up to a factor of three should be observable in 
$~^{41}\text{K}$/${}^{87}\text{Rb}$ systems. Note the particle mass renormalization is stronger for lighter impurities.

%============================================================

%============================================================

{\it Experimental signatures of correlations}. The main new features of the 
CGW (\ref{CGW}) compared to the  mean-field wavefuction (\ref{mean_field}) are correlations
between different phonon modes. Such correlations will also be present for atoms of the host BEC itself and
can be measured using noise correlation analysis in the time-of-flight experiments (TOF)
~\cite{bloch_bosons,bloch_fermions}. The quantity that can be extracted from TOF images is
the second order correlation function~\cite{altman}
\begin{equation}
g^{(2)}(k,k')\approx \frac{\av{\ad_k \ad_{k'} \a_k \a_{k'}}}{\av{\ad_k \a_k}\av{\ad_{k'} \a_{k'}}}.
\label{Eq:g2}
\end{equation}
Note that we will focus on the additional correlations among host atoms caused by the impurity and will not include correlations present in the BEC itself. The latter are expected only for $\pm k$ atoms, as described by the Bogoliubov wavefunction
~\cite{Westbrook_experiment}. 
Fig.\ref{fig:2} presents results of correlations described by the equation (\ref{Eq:g2}) for the experimental
systems $^{41}$K/$^{87}$Rb and $^{133}$Cs/$^{87}$Rb. 
It is possible to obtain asymptotic values of these correlations in several regimes. 
In the long wavelength limit phonon modes decouple and $g^{(2)}$ approaches unity.
For high momenta occupation numbers of atoms $n_k$ decrease but $g^{(2)}(k,k')$ saturates at values that
depend on the angle $\theta$ between $\vec{k}$ and $\vec{k}'$: $ g^{(2)}(k=\infty, k'=\infty)\left( 1+\sqrt{2}m_B/M\right)^{-1} $
 and $g^{(2)}(k=\infty, k'=-\infty)=\left( 1+\sqrt{2}m_B/M\right)$. This indicates
antibunching of bosons for small $\theta$ and bunching for $\theta = \pi$. 
These results are consistent with our intuition that an impurity colliding with one of the BEC 
atoms and giving it momentum  $\vec{k}$  is more likely to scatter the next BEC atom in the opposite 
direction.
Correlations induced between host atoms are stronger for light impurities. 
One of the intriguing features in Fig.\ref{fig:2}(b) is a peak in the correlation
function at stronger coupling $\alpha=4$ and $k\xi \approx 2$.    

%============================================================
Before concluding this paper we point out that wavefunctions~\eqref{CGW} are commonly used in quantum optics~\cite{Loudon1987,Zhang1990,Loudon2000,Dodonov2002,Walls2007}.
However theoretical analysis so far focused either on time dependent quadratic Hamiltonians, non-linear Hamiltonians
with only few modes, for which direct optimization is possible, or many body multimode Hamiltonians
that have translational symmetry, which allowed factorization the many body wavefunctions into separate
contribution from $\left( k, -k\right)$ pairs (translational invariance allows only $\av{b_k^\dag b_{-k}}$ and $\av{b_k^\dag b_k}$)
expectation values). We expect that the approach developed in this paper can become a useful tool for analyzing quantum optical systems
with many modes, strong nonlinearities, and no translational symmetry, such as Rydbergs systems,
circuit QED, coupled non-linear resonators, and plasmonic systems.

%============================================================

{\it Summary and outlook}. We proposed a class of variational Gaussian wavefunctions for Fr\"ohlich polarons that gives 
excellent agreement with Monte-Carlo results for the polaron energy 
in a wide range of parameters. We find a smooth crossover of the effective polaron mass as the interaction strength changes from from weak to intermediate coupling.
Our wavefunction predicts a specific pattern of correlations between host atoms that
can be measured in TOF experiments. We suggest that our predictions can be checked in such systems as
$^{41}$K or  $^{133}$Cs impurities in $^{87}$Rb BEC, in which the dimensionless coupling constant can reach the value as large as 4,
while the  Fr\"ohlich polaron description remains appropriate. 
We point out that Gaussian wavefunctions can be used to describe not only equilibrium states but also dynamics. 
Thus our formalism can be extended to compute spectral functions of polarons and study response of polarons to external fields.

%============================================================

We acknowledge useful discussions with 
D. Abanin,
I. Bloch, 
M. Fleischhauer,
T. Giamarchi, 
S. Giorgini,
W. Hofstetter, 
L. Keldysh, 
M. Oberthaler, 
D. Pekker, 
A. Polkovnikov, 
L. Pollet, 
N. Prokof’ev, 
S. Das Sarma, 
L. Pitaevskii,
R. Schmidt, 
A. Shashi,
V. Stojanovic, 
L. Tarruell, 
N. Trivedi, 
A. Widera and 
M. Zwierlein. 
We also thank W. Casteels for providing us with results for the polaron mass obtained by Feynman's variational approach presented in Fig~\ref{fig:Mass_theory_comparison}.
The authors acknowledge support from the NSF grant DMR-1308435, Harvard-MIT CUA, AFOSR New Quantum Phases of Matter MURI, the ARO-MURI on Atomtronics, ARO MURI Quism program.
Y.E.S. and A.N.R. thank the Dynasty foundation and the Russian Foundation for Basic Research grant 14-02-01219 for financial support. 
F.G. is a recipient of a fellowship through the Excellence Initiative (DFG/GSC 266) and is grateful for financial support from the ”Marion K\"oser Stiftung”. 

%============================================================

\bibliography{Polaron}

\newpage

%%%%%%%%%%%%%%%%%%%%%%%%%%%%%%%%%%%%%%%%%%%%%%%%%%%%%%%%%%%%%%%%%%%%%%%%%%%%%%%%%%%%%%%%%%%%%%%%%%%%%%%%%%
%%%%%%%%%%%%%%%%%%%%%%%%%%%%%%%%%%%%%%%%%%%%%%%%%%%%%%%%%%%%%%%%%%%%%%%%%%%%%%%%%%%%%%%%%%%%%%%%%%%%%%%%%%

\begin{widetext}
\begin{center} { \large \bf Supplementary Materials for "Polaronic mass renormalization of impurities in BEC: correlated Gaussian wavefunction approach"} \end{center}

%%%%%%%%%%%%%%%%%%%%%%%%%%%%%%%%%%%%%%%%%%%%%%%%%%%%%%%
% Counter resetting for supplementaries

\setcounter{figure}{0}   \renewcommand{\thefigure}{S\arabic{figure}}

\setcounter{equation}{0} \renewcommand{\theequation}{S.\arabic{equation}}

\setcounter{section}{0} \renewcommand{\thesection}{S.\Roman{section}}

\renewcommand{\thesubsection}{S.\Roman{section}.\Alph{subsection}}

\makeatletter
\renewcommand*{\p@subsection}{}  
\makeatother

\renewcommand{\thesubsubsection}{S.\Roman{section}.\Alph{subsection}-\arabic{subsubsection}}

\makeatletter
\renewcommand*{\p@subsubsection}{}  % referring to subsubsections
\makeatother

%%%%%%%%%%%%%%%%%%%%%%%%%%%%%%%%%%%%%%%%%%%%%%%%%%%%%%%

%%%%%%%%%%%%%%%%%%%%%%%%%%%%%%%%%%%%%%%%%%%%%%%%%%%%%%%%%%%%%%%

In this supplementary materials we present details of the calculation of the polaron energy using CGW given in eq.~(5) of the main text.
We then show how one can minimize this energy to obtain self-consistency equations for the variational parameters. 

\paragraph{Variational Gaussian Approach.}
Our starting point is the Fr\"ohlich Hamiltonian after the Lee Low Pines~\cite{Lee1953} transformation
\begin{equation}\label{eq:H}
\av{\hat{\mathcal{H}}_{\rm LLP}}=\frac{1}{2M}\left(\v P-\sum_{k}\v{k}\hat{b}_{k}^{\dagger}\hat{b}_{k}\right)^{2}
+\sum_{k}V_{k}\left(\hat{b}_{k}^{\dagger}+\hat{b}_{-k}\right)+\sum_{k}\omega_{k}\hat{b}_{k}^{\dagger}\hat{b}_{k}.
\end{equation}

Gaussian wavefunctions take into account entanglement between different phonon modes, which are absent in mean field theories. 
As a consequence pairwise averages, e.g., $\av{b_k b_{k'}}$, have a nonzero irreducible part.
Because of the Gaussian statistics all higher-order correlators as $\av{b_k^\dag b_{k'}^\dag b_k b_{k'}}$ 
can be reduced to simple two-point expressions using Wick's theorem.
In particular the average of Fr\"ohlich Hamiltonian over arbitrary Gaussian trial state $\av{\hat{\mathcal{H}}_{\rm LLP}}$ becomes

\begin{multline}\label{eq:GaussHamilt}
\av{\hat{\mathcal{H}}_{\rm LLP}}=\frac{P^2}{2M}+\frac{1}{\sqrt{V}}\sum_{k}V_{k}\left(\av{\hat{b}_{k}^{\dagger}}+\av{\hat{b}_{k}}\right)+
\sum_{k}\left( \omega_{k}+\frac{k^2}{2M}-\frac{\v P \v k}{M}+\frac{\v k}{2M}\sum_{k'} \v k' \av{\hat{b}_{k'}^{\dagger}\hat{b}_{k'}}\right)\av{\hat{b}_{k}^{\dagger}\hat{b}_{k}}+\\
\frac{1}{2M}\sum_{kk'}\v k \v k'\left( 
\av{\hat{b}_{k}^{\dagger}}\av{\hat{b}_{k'}^\dag}\av{\hat{b}_{k}^{}\hat{b}_{k'}}_c +
\av{\hat{b}_{k}^{\dag}\hat{b}_{k'}^\dag}_c\av{\hat{b}_{k}^{}}\av{\hat{b}_{k'}}+
\av{\hat{b}_{k}^{\dagger}}\av{\hat{b}_{k'}}\av{\hat{b}_{k'}^{\dagger}\hat{b}_{k}}_c+
\av{\hat{b}_{k}^{\dagger}\hat{b}_{k'}}_c\av{\hat{b}_{k'}^{\dagger}}\av{\hat{b}_{k}}\right)+\\
\frac{1}{2M}\sum_{kk'}\v k \v k'\left( 
\av{\hat{b}_{k}^{\dag}\hat{b}_{k'}^\dag}_c \av{\hat{b}_{k}^{}\hat{b}_{k'}}_c+
\av{\hat{b}_{k}^{\dagger}\hat{b}_{k'}}_c\av{\hat{b}_{k'}^{\dagger}\hat{b}_{k}}_c\right).
\end{multline}

where we defined the irreducible connected correlations as $\av{\hat A \hat B }_c = \av{\hat A\hat B}- \av{\hat A} \av{\hat B}$.

Our variational CGW given by eq.~(5) of the main text give the most general Gaussian wavefunctions.
For the ground state (equilibrium) problem under consideration it is sufficient to consider real vector $\beta$ and real symmetric matrix $Q$,
up to an overall phase, which provide minimum to the energy in eq.~(\ref{eq:GaussHamilt}).

The unitary transformations $\hat D(\lbrace\beta \rbrace) \hat S(\lbrace Q\rbrace)$ can be understood either as a transformation of the bosonic vacuum wavefunction 
into a correlated Gaussian state $\ket 0 \to \ket{\text{CGW}}$, or as a Bogoliubov rotations
of the creation  (annihilation) operators. 
To evaluate $\av{\hat{\mathcal{H}}_{\rm LLP}}$ in Eq. (2) with the CGWs, we find it most convenient to perform a Bogoliubov basis transformation,
\begin{equation}\label{Bogoluibov}
\hat{\mathcal{B}}_k \equiv \hat S^\dag(\lbrace Q\rbrace)  \hat D^\dag(\lbrace\beta\rbrace) \hat b_k \hat  D(\lbrace\beta\rbrace) \hat S(\lbrace Q\rbrace) = 
\beta_k + \sum_{k'} [\cosh Q]_{kk'} \hat b_{k'} +\sum_{k'} [\sinh Q]_{kk'} \hat b^\dag _{k'}
\end{equation}
and calculate the vacuum expectation value in the new basis, e.g.  
$\av{\hat{\mathcal{ H}}_{LLP}\left( \hat b_k^\dag,\hat b_k\right)}=\bra{0} \hat{\mathcal{H}}_{LLP}\left( \hat{\mathcal{B}}_k^\dag,\hat{\mathcal{B}}_k\right) \ket{0}$.
Here, and in what follows, functions of the matrix Q (e.g. $\cosh Q$) should be understood as being defined through their Taylor expansion.
Using the relation ~(\ref{Bogoluibov}), we can now calculate the irreducible two-point functions required to evaluate the variational energy,
\begin{eqnarray}\label{eq:averages}
\av{\hat b_k} = \beta_k, \;\;\; \av{\hat b_k \hat b_{k'}^\dag }_c=\frac{1}{2}\left[\cosh 2Q\right]_{kk'},\;\;\; \av{\hat b_k \hat b_{k'} }_c = \frac{1}{2}\left[\sinh 2Q\right]_{kk'}
\end{eqnarray}

In order to derive self-consistency equations for $\beta$ and $Q$, we minimize the variational energy ~(\ref{eq:GaussHamilt}) 
with the expectation values given by eq.~(\ref{eq:averages}). In addition, to obtain tractable equations, 
we consider only terms up to second order in $Q$ in the energy $\av{\hat{\mathcal{H}}_{\rm LLP}}$. 
Physically, this corresponds to the assumption that phonon-phonon correlations are small, albeit non-vanishing. 
Note that this truncation can not be justified on the ground that matrix elements $Q_{kk'}$ are of the order of inverse volume $1/\mathcal{V}$. 
Summations implied in matrix multiplication $\left[Q^2\right]_{kk'}= \sum_p Q_{kp}Q_{pk'} = \mathcal{V} \int_p Q_{kp} Q_{pk'} $ show that higher order terms have the same scaling in powers of $1/\mathcal{V}$. 
However analysis shows that even for intermediate interaction strength the matrix norm $\Vert Q \Vert$ is numerically small justifying the expansion.
Thus we obtain the truncated variational energy

\begin{multline}\label{eq:AvEnApprox}
\av{H}=\frac{P^2}{2M}+2\sum_k~V_{k}\beta_k+
\sum_k~ \left( \omega_k +\frac{k^2}{2M} -\frac{\v P \v k}{M}\right) \left( \beta_k^2+\sum_{k'}~Q_{kk'}^2\right)+
\sum_{kk'}~\frac{\v k \v k'}{2M}\beta_k^2 \beta_{k'}^2+\\
\sum_{kk'q}~\frac{\v k \v k'}{M}\beta_k^2 Q_{k'q}^2+
\sum_{kk'}~\frac{\v k \v k'}{M}\beta_k \beta_{k'}Q_{kk'}+
\sum_{kk'q}~\frac{\v k \v k'}{M}\beta_k \beta_{k'}Q_{kq}Q_{qk'}+
\sum_{kk'}~\frac{\v k \v k'}{2M}  Q_{kk'}^2
\end{multline}
To find the minimum of~(\ref{eq:AvEnApprox}) we vary the last expression with respect to $\beta$ and $Q$, and derive the self-consistency equations.

\paragraph{Equations for $Q_{kk'}$.}
Minimization  with respect to $Q$ gives
\begin{equation} 
\left(\Omega_{k}+\frac{\sc{k}{k'}}{M}+\Omega_{k'}\right)Q_{kk'}+
\frac{\sc{k}{k'}}{M}\beta_k \beta_{k'}+
\sum_q~\frac{\sc{q}{k'}}{M}\beta_q \beta_{k'}Q_{kq}+
\sum_q~\frac{\sc{k}{q}}{M}\beta_q \beta_{k}Q_{qk'}=0
\label{eq:Q}
\end{equation}
where the dispersion relation reads
\begin{equation}
\Omega_k= \omega_k +\frac{k^2}{2M} -\frac{\v P\v k}{M}+\frac{\v k}{M}\sum_{k'} \v k' \beta_{k'}^2.
\end{equation}
It is similar to the mean field expression (see eq. (3) in the main text), except that the polarization 
$\beta_k$ is now determined by a different self-consistent procedure.

To cast Eq.~(\ref{eq:Q}) into a more tractable form, we now define the following auxiliary quantities,
$\eta_{k,k'}$ and $D(k,k')$
\footnote{The quantity $D(k,k')$ is in fact the two particle excitation energy over the vacuum state $\left|0\right.>$. 
One can check this using the standard 4-th order
perturbation  theory with respect to interaction $V_k$.} 
by the following formulas
\begin{eqnarray}\label{eq:Qeta}
\eta_{k,k'} &=& - M Q_{kk'} \frac{D(k,k') }{\beta_k\beta_{k'}}, \\ \nonumber
D(k, k')&=&\Omega_{k}+\frac{\sc{k}{k'}}{M}+\Omega_{k'}.
\end{eqnarray}

We express  $Q_{kk'}$ via $\eta_{kk'}$ and substitute it into the equation~(\ref{eq:Q}):
\begin{equation}\label{eq:eta}
\eta_{k,k'}=
\sc{k}{k'}-
\sum_q~ \frac{\beta_{q}^2}{M} \left( \frac{\eta_{k,q}\sc{q}{k'} }{D(k,q)}   +
\frac{\sc{k}{q}  \eta_{q,k'} }{D(q,k')} \right).
\end{equation}
Let us now introduce the vector
\begin{equation}
\v F_k = \sum_q~ \frac{\beta_{q}^2}{MD(k,q)} \eta_{k,q} \v q,
\end{equation}
so that  the equation~(\ref{eq:Q}) takes a particularly simple form
\begin{eqnarray} \label{eq:eta}
\eta_{k,k'}&=&\v k \v k'-\v{F}_k\v{k}' -\v k \v F_{k'}.
\end{eqnarray}
We introduce the tensorial quantities $\tens{A}$ 
\begin{eqnarray}
\tens{A}_k^{(0)} &=&\sum_{k'} \frac{\beta_{k'}^2}{MD(k,k')} \v k' \otimes  \v k' ,\\
\tens{A}_k^{(1)} &=&\sum_{k'} \frac{\beta_{k'}^2}{MD(k,k')} \v  F_{k'} \otimes \v k' ,\\
\tens{A}_k^{(2)} &=&\sum_{k'} \frac{\beta_{k'}^2}{MD(k,k')} \v F_ {k'}\otimes \v F_{k'}.
\end{eqnarray}
where the outer product of two vectors is $\v{k} \otimes \v{k}'$.

\begin{figure}[t]
\includegraphics[width=0.25\columnwidth]{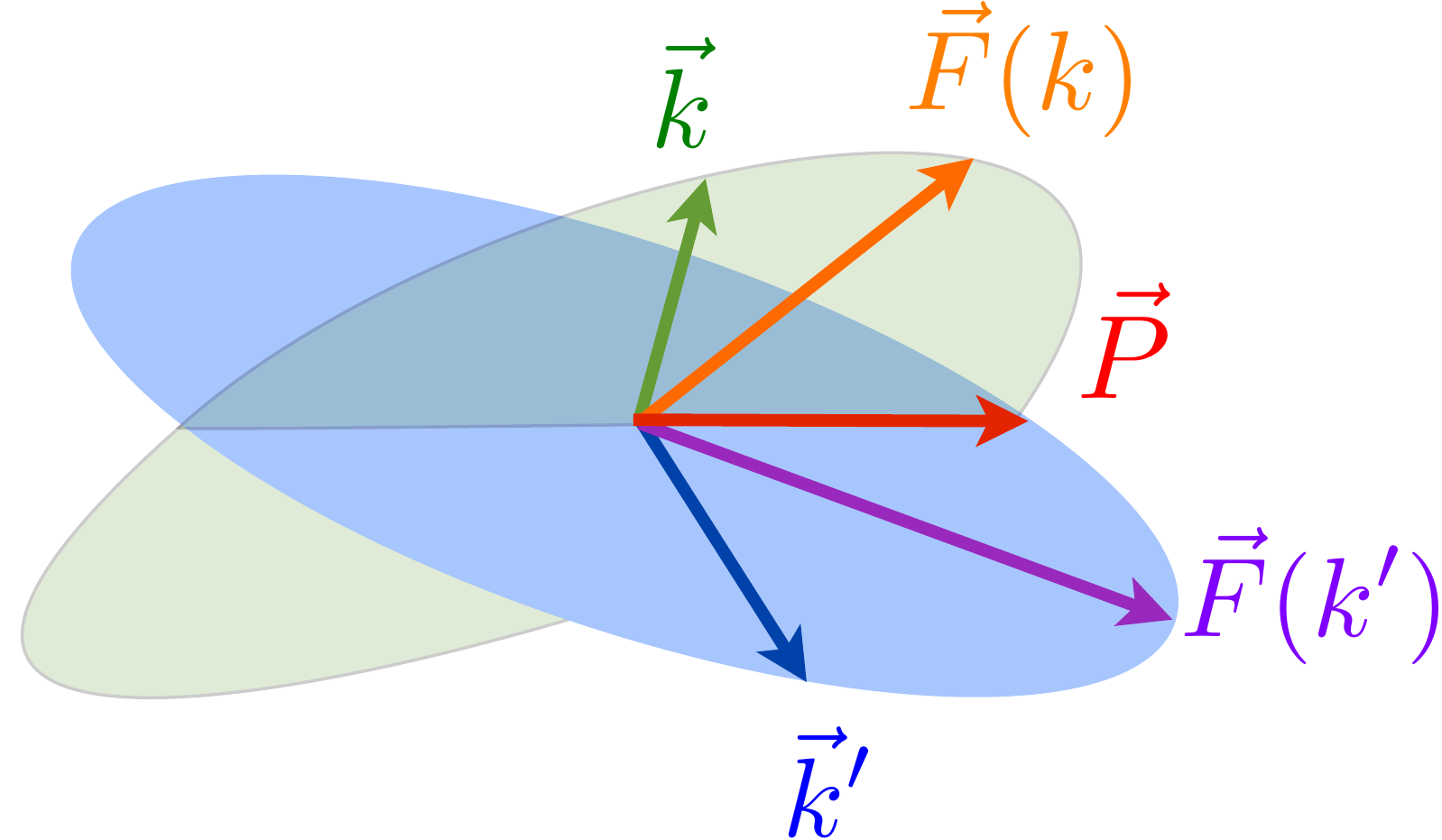} 
\caption{Schematic of the respect direction of vectors: total momentum $\v P$, momentum of given mode $\v k$,
and vector $\v F(k)$. Vectors $\v P$ and $\v k$ define a plane $\left(\v P, \v k \right)$ and vector $\v F (k)$ is in this plane.}
\label{FIG:F}
\end{figure}

Then a multiplication of the equation~(\ref{eq:eta}) by $\frac{\beta_{k'}^2}{MD(k,k')} \v k'$ and subsequent summation
 over $k'$ gives the self-consistency equation for the vector $\v F_k$
\begin{equation} \label{eq:F}
\v F_k=\left( \v k - \v{F}_k  \right) \tens{A}_k^{(0)}
-\v k~ \tens{A}^{(1)}_k.
\end{equation}
This equation is solved numerically together with the equation for $\beta$, which will
be derived in the next subsection.

Let us first discuss the geometrical properties of the vector $\v F_k$. In case of $P=0$ vector
 $\v F_k$ is collinear to  $\v k$ as there are no other vector quantities in the formalism.
Formally this corresponds to $\v F_k  = R_k \v k$, with the proportionality coefficient $R_k$.
In general case $P \neq 0$ 
$\v F_k $ belongs to the plane of the vectors $\v P$ and $\v k$, and 
$\tens{R}_k$ is a tensor that describes a combination of rotation in this plane and rescaling ( see Fig.\ref{FIG:F} for illustration).

\paragraph{Equations for $\beta_k$.}

Variation of the expression~(\ref{eq:AvEnApprox}) with respect to $\beta$ gives 
\begin{equation}\label{eq:a0}
V_{k}+\beta_k\left( \omega_k +\frac{k^2}{2M} -\frac{\v k}{M}\left( \v P -\v P_\mathrm{ph}\right)\right)+
\frac{\beta_k \v k}{M}\sum_q~\frac{\beta_q^2 }{M D(k,q)  }\eta_{q,k} \left( \v F(q) -\v q\right)=0
\end{equation}
where we substituted $Q$ with the corresponding expressions~(\ref{eq:Qeta}) in terms of $\eta_{k,k'}$ after the variation.
The total momentum carried by the phonons is $\v P_\mathrm{ph} = \sum_{kk'} \v k \left( \beta_k^2 \delta_{kk'}+ Q_{kk'}^2\right)$.
The last term on the left hand side of Eq.~(\ref{eq:a0}) can be interpreted as a renormalizition of the phonon dispersion relation $\Omega_k$.
Let us rewrite the expressions so that this statement is more clear.
We use equations~(\ref{eq:eta}) and~(\ref{eq:F}), and also recall the geometrical properties of vector $\v F_k$ 
discussed above: $\v F_k  =  \tens{R}_k \v k$. We rewrite the expression~(\ref{eq:a0}) as follows
\begin{equation}\label{eq:a1}
V_{k}+\beta_k\left( \omega_k +\frac{k^2}{2M} -\frac{\v k}{M}\left( \v P -\v P_\mathrm{ph}\right)+\right.\\\left.
\v k\frac{1}{M}\left( \left( \tens{A}^{(1)}_k - \tens{A}^{(0)}_k\right)\left(  \tens{I}- \tens{R}_k\right) -\tens{A}^{(2)}_k \right)\v k\right)=0.
\end{equation}
Thus equation for $\beta_k$ can be written in a compact form
\begin{equation}\label{eq:beta}
\beta_k=-\frac{V_k}{ \omega(k)+\frac{ \v k \tens{\mathcal{M}}^{-1}(k) \v k}{2} +\frac{\v k}{M} \left( \v P - \v P_B\right)}.
\end{equation}
Here the effective impurity mass
\begin{equation}
M \tens{\mathcal{M}}^{-1}(k) = \tens{I}- 2
\left( \left( \tens{A}^{(1)}_k - \tens{A}^{(0)}_k\right)\left(  \tens{I}- \tens{R}_k\right) -\tens{A}^{(2)}_k \right)
\end{equation}
is a tensor quantity which is non-diagonal for $P\neq 0$.

\paragraph{Observables.}
Equations~(\ref{eq:F}) and~(\ref{eq:beta}) for $\beta_k$ and $\v F_k$ form a self-consistent set for $\beta_k$ and $\v F_k$ which we solve iteratively.
After obtaining $\beta_k$ and $\v F_k$ all observables can be calculated using Wick's theorem. In particular the polaron energy reads
\begin{equation}
E_p=\av{\hat{\mathcal{H}}_{\rm LLP}} = \frac{\v P^2}{2M} - \frac{\v P_\text{ph}^2}{2M} +\sum_k V_k \beta_k -\sum_k  \beta_k^2\frac{\v F^2_k}{2M}
+\sum_k  \beta_k^2 \frac{k_\mu \left( M^{-1} \delta_{\mu \nu}  - \mI_{\mu \nu}(k)\right) k_{\nu}}{2}
\end{equation}
It is important to point out that only the total impurity energy $E=g_{\rm IB} n_0+E_p$ is physically meaningful, 
where the first term is the bare impurity-condensate interaction energy. 
The energy $E_p$ by itself is UV-divergent. This divergence appears at the mean field level
and comes from the term $\sum_k V_k \beta_k$. Indeed one can check that in UV-limit  $\beta_k\propto k^{-2}$ and $V_k$ tends to a constant value.
Therefore in $d>2$ this gives rise to a power-law divergency of the polaron energy $\sum_k V_k \beta_k \propto \Lambda^{d-2}$, 
where $\Lambda$ is a sharp UV momentum cut-off. 
This divergence is resolved by the standard regularization procedure~
\cite{Tempere2009,aditya,Rath2013}, expressing $g_{IB}$ in terms of the scattering length $a_{IB}$ and the cut-off $\Lambda$.
When quantum fluctuations are taken into account and an additional logarithmic divergence with $\Lambda$ appears as we discuss in detail in~\cite{RG}.
The presence of this logarithmic behavior makes a direct comparison with the experimental data involved.
Thus our results for polaron energy are only used to benchmark the approach by comparing to other known theoretical results. 

\begin{figure}[t]
\includegraphics[width=1\columnwidth]{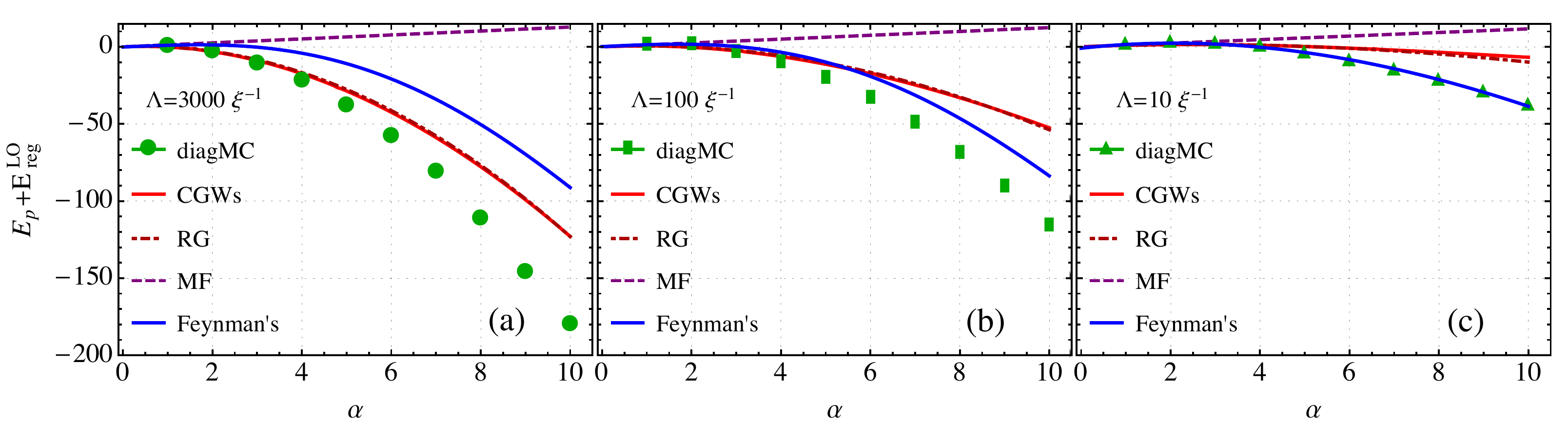} 
\caption{ {Polaron energy $E_p + E_\text{reg}^\text{LO}$ for static $^6$Li impurity ($P=0$) in $^{23}$Na BEC predicted by different theoretical approaches
as a function of the dimensionless coupling constant $\alpha$ in the strong coupling regime for different values of cut-off parameter: 
(a) $\Lambda = 3000 \; \xi^{-1}$,  (b) $\Lambda =100 \; \xi^{-1}$,  (c) $\Lambda = 10 \; \xi^{-1}$.
Our result (CGWs) is compared with MC calculations~\cite{Vlietinck2015}, Feynman's variational method~\cite{Tempere2009}, mean-field~\cite{aditya}, and Renormalization group~\cite{RG}.}}
\label{FIG:F}
\end{figure}

\begin{figure}[t]
\includegraphics[width=0.4\columnwidth]{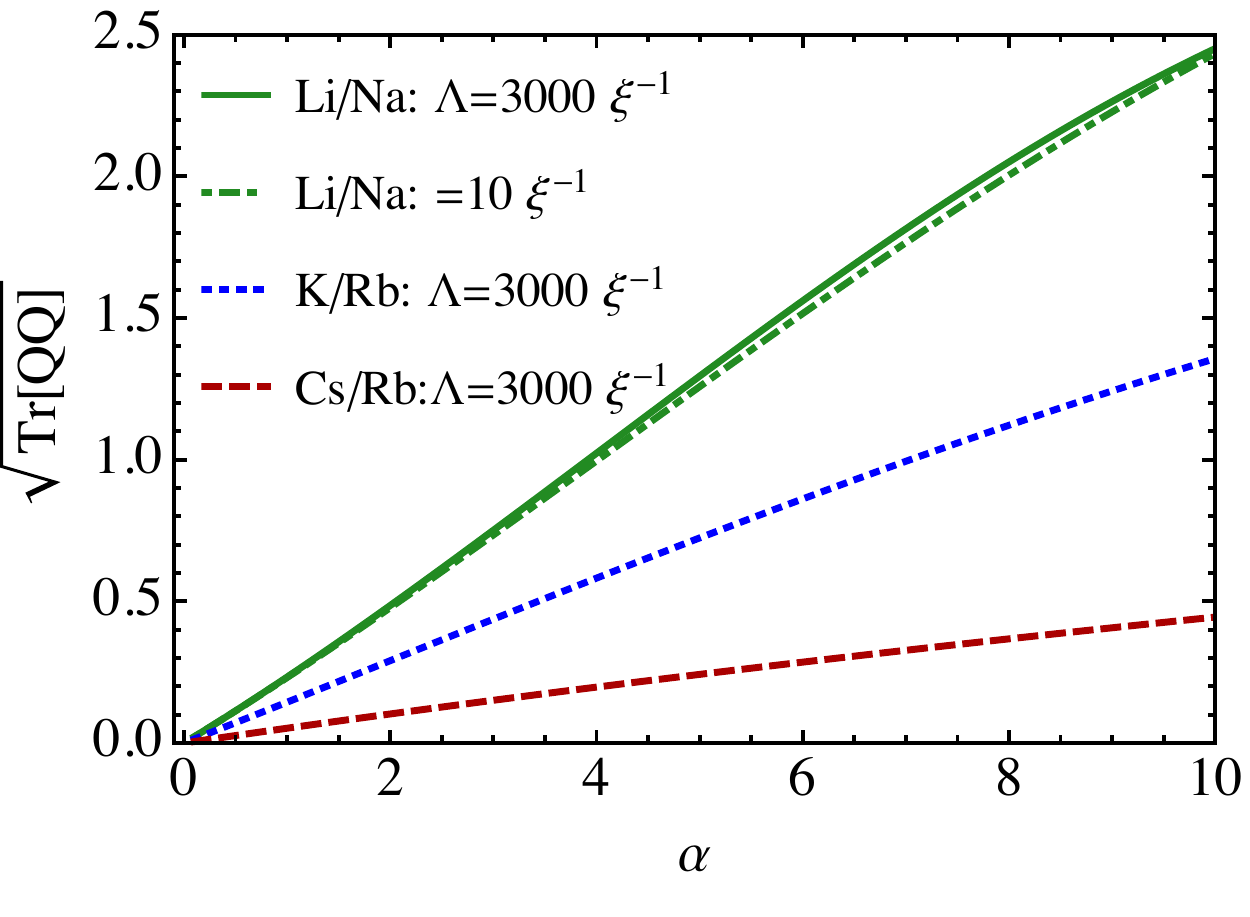} 
\caption{ {Measure of correlation strength, $\sqrt{Tr[QQ]}$, as a function of the dimensionless coupling constant $\alpha$ for systems with various impurity and host species: Li/Na ($M/m_B =0.23$), K/Rb ($M/m_B =0.47$), and Cs/Rb ($M/m_B =1.53$). 
%Note that $\sqrt{Tr[QQ]}$ strongly depends on the mass ratio and is independent on the UV cut-off $\Lambda$.}
}}
\label{FIG:2}
\end{figure} 

 {
The ground-state energy of the Fr\"ohlich Hamiltonian with the regularized leading order divergence $E_p+E_\text{reg}^{LO}$
is shown in Fig. 2 in the main text and in Fig.~\ref{FIG:2}.
The results obtained by the CGWs approach is in a good agreement with the diagMC up to $\alpha=4$ for any value of the UV cut-off parameter. In the strong coupling limit there is a discrepancy between numerically exact solution and the results obtained by the CGWs, 
which is due to the perturbative expansion of the energy as a function of the squeezing parameter $Q$. 
To supplement this statement with concrete numbers we calculate the matrix norm of the squeezing parameter $\sqrt{Tr[QQ]}$, shown in Fig. \ref{FIG:2}. 
The perturbative expansion is no longer valid when the matrix norm is of the order of unity, $\sqrt{Tr[QQ]} \approx 1$. Note that the squeezing parameter is smaller for heavier impurities, since the nonlinear term in the Hamiltonian~\eqref{eq:GaussHamilt} is proportional to the inverse mass of the impurity $M^{-1}$. 
}

\paragraph{Noise correlations.} 
In the main text of the paper we showed that impurity atoms create additional correlations among host atoms, which we describe using Gaussian variational wavefunctions. A formal way of describing these correlations is via the second order correlation function $g^{(2)}(k,k')$.
We point out that while our analysis considered only a single impurity, experiments are performed at finite impurity concentration. Assuming that impurities are sufficiently dilute and their polarization clouds do not overlap, we can neglect interaction between polarons. Then changes in the occupation number of
host bosons at finite $k$  due to several impurities will be proportional to the number of impurities. 
In the case of $^{41}$K/$^{87}$Rb mixtures, interaction strength $\alpha=4$, and impurity concentration 5 per cent, 
we estimate the number of atoms excited from the condensate to finite momentum states by scattering on impurity atoms to be 3 per cent. 
This sets the magnitude of the signal whose correlations we discuss in the main text.

\end{widetext}

\end{document}